\documentclass[10pt]{iopart}
\usepackage{mathptmx}
\usepackage{amssymb}
\usepackage{bm}
\usepackage{iopams}
\usepackage{color}
\definecolor{colorLink}{rgb}{0.6,0,0}
\definecolor{colorCite}{rgb}{0,0,1}
\definecolor{colorURL}{rgb}{0,0.6,0.0}
\usepackage[pdftitle={s0},colorlinks=true,linktocpage=true,linkcolor=black,citecolor=colorCite,urlcolor=colorURL]{hyperref}

\usepackage{booktabs}
\usepackage{graphicx}
\usepackage{multirow}
\usepackage[numbers,sort&compress]{natbib}
\usepackage{url}

\begin{document}
	\Huge
	\title{Uni-layer magnets: a new concept for LTS and HTS based superconducting magnets}
	
	\author{José Luis Rudeiros Fernández and Paolo Ferracin}
	
	\address{Lawrence Berkeley National Laboratory, Berkeley, CA 94720, USA}
	\ead{jrudeirosfernandez@lbl.gov}
	\vspace{10pt}
	\begin{indented}
		\item[]December 2022
	\end{indented}
	
	\begin{abstract}
		\normalsize
		A novel geometrical configuration to form a magnetic field perpendicular to an aperture, created by an asymmetric current distribution, within a single layer, and using a continuous ideal current line, named the \textit{uni-layer} magnet, is here presented. The idea is compared to existing concepts in superconducting magnets, namely, the $\cos{\theta}$ sector magnet, stress managed $\cos{\theta}$ and canted $\cos{\theta}$. The \textit{uni-layer} magnet allows for a design with a continuous unit length (no layer jump), and an increased minimum bending radius of the conductor in relation to traditional $\cos{\theta}$ and canted $\cos{\theta}$ designs. The specific characteristics of the \textit{uni-layer} design are especially advantageous for strain-sensitive and prone to winding degradation high-temperature superconductors, in very high field accelerator magnet applications, in which, high efficiency in the use of conductor, and a small aperture are required. The advantages with regard to the design and fabrication of \textit{uni-layer} magnets in relation to other concepts are also discussed.
	\end{abstract}
	\ioptwocol
	\normalsize
		\section{Introduction}
	There are multiple spatial configurations in which to arrange a set of current lines (or current density) to form a desired magnetic field that is perpendicular to the longitudinal direction of a free region of space, i.e. aperture. This condition of perpendicularity is required in accelerator magnets, for which the magnetic field shall be perpendicular to the beam direction, i.e. moving along the aperture.
	
	From idealized current distribution cases, \textit{real world} approximations are normally derived due to the constraints imposed by conductors with finite and constant cross sections. This is the case for the most extended coil layouts used for accelerator magnets, such as the $\cos{\theta}$ sector coil magnet (CT) \cite{Koepke1979,Wolff1988,Greene1996,Rossi2004,Xu2013} (related to the idea of a $\cos{\theta}$ current distribution), the block type \cite{Taylor1985,Sabbi2005,Milanese2012,RudeirosFernandez2022} and common coil magnets \cite{Danby1983,Gupta1998,Toral2018}  (related to the idea of an infinite slab dipole), and canted $\cos{\theta}$ (CCT) \cite{Meyer1970,Caspi2007,Brouwer2015} (related to the idea of a solenoid).
	
	In the case of very high field magnets \cite{McIntyre2005,Ferracin2022,Ferracin2023} using strain sensitive conductors, the electromagnetic forces acting in the conductor lead to the need for \textit{stress-managed} concepts, in which part of the force is captured by a structural element of the magnet, reducing the overall accumulated force over the coil's cross-section area, and therefore limiting the stress and strain applied to the conductor during energization of the magnet. For this reason, not only is the efficiency of how a magnetic field is formed by certain current distribution relevant, but also the ability of the magnet design to create the framework for the conductor to operate under extreme conditions. In this regard, the CCT as conceived in \cite{Caspi2007,Brouwer2015}, is intrinsically a stress-managed concept, where the structure, formed by the spar and ribs, intercept part of the Lorentz force of each turn. In the case of CT magnets, various designs for stress-managed CT (SMCT) \cite{Patoux1983,Zlobin2018,Ferracin2022,Ferracin2023} have also been proposed. Although SMCT magnets can accommodate the interception of block segments, it seems that within the inner layers of very high field magnets, such as the $20\,\textrm{T}$ magnet discussed in \cite{Ferracin2023}, where the use of high temperature superconductors (HTS) is necessary, the support of each individual turn is required in order to maintain the strain-stress of the conductor to acceptable levels.
	
	In addition to the very high forces imposed in the conductor, when using HTS such as REBCO, and derived wires such as {\sc corc}\textsuperscript{\textregistered} \cite{Weiss2017,VanDerLaan2019}, one of the main difficulties is their susceptibility to \textit{degradation} when wound around relatively small apertures \cite{Wang2021,Stern2022}. As the use of HTS is normally more advantageous in the high field region of the magnet, the minimum radius of curvature required by the winding is relatively low. This issue leads to the impossibility of using this type of conductor in a traditional CT configuration, and leading to the use of less efficient designs options (e.g. CCT with relatively high inclination \cite{Wang2021}), that can lead a significantly higher overall cost.
	
	In the spirit of addressing the previous issues, this paper presents a new concept to form a magnetic field that is perpendicular to an aperture: the \textit{uni-layer} magnet (UL). The idea is derived from the harmonic expansion of the magnetic field, and the concept of creating \textit{quasi-perfect} fields even in partially asymmetric geometrical configurations. In this regard, we will demonstrate the existence of asymmetric configurations that satisfy the field quality specifications normally required in accelerator magnets. We will also demonstrate that the partial asymmetry allows for a new spatial configuration of the conductor around the aperture, in which a single layer with no internal layer jump is possible, facilitating therefore magnet layer \textit{grading}. Furthermore, we will also demonstrate that these asymmetric solutions lead to a significant increase in the minimum radius required to wind a conductor around a specific aperture in relation to CT and CCT magnets. The UL concept is then discussed in relation to the CT/SMCT and CCT, showing how it presents several advantages in terms of design, manufacturing, and performance for low temperature superconductor (LTS), and especially HTS applications in very high field magnets for high energy accelerators.	
	
	\section{Uni-layer magnets}
	
	An idealized \textit{uni-layer} magnet can be defined as a magnet that generates a $\mathbf{B}$ field within a straight region of space, by a system of ideal current lines, all parallel to the \textit{z-axis} (along the straight section of the magnet) of a Cartesian coordinate system (i.e. perpendicular to the $xy$-plane), that lay within a single continuous surface (i.e. within a single layer), and that are connected by a single continuous path that does not cross itself. In this regard, a magnet could also be constituted by a set of individual uni-layers.
	
	Let's also define two main categories of UL magnets: \textit{symmetric }and \textit{asymmetric}. A \textit{symmetric} UL magnet will contain the same number of current lines when angularly circulating between poles. An \textit{asymmetric} uni-layer magnet will contain a non-equal number of current lines when circulating between poles. A special case of an asymmetric UL, which will be examined in the next section, will lead to the possibility of a winding configuration within a single layer of zero Gaussian curvature.
	
	The search for the position and current magnitude of the idealized current lines to form a specific $\mathbf{B}_{objective}$ within the \textit{straight section} of the magnet can be treated as a global optimization problem. For a number of $m$ current lines, we can define an array containing the argument 
		\begin{equation*}
		\label{deqn_theta_array}
		\theta = \left\{  \theta_1,\theta_2,\dots \theta_m \right\}
	\end{equation*}
	and modulus
	\begin{equation*}
		\label{deqn_modulus}
		\rho = \left\{  \rho_1,\rho_2,\dots \rho_m \right\}
	\end{equation*}
	of the current lines in the complex plane, as well as the current
	\begin{equation*}
		\label{deqn_current}
		I = \left\{  I_1,I_2,\dots I_m \right\}
	\end{equation*}
	leading to a general set of equations that can be expressed as

	\begin{equation}
			\label{deqn_opti_all}
		\eqalign{\textrm{minimize} \|\mathbf{B}_{objective} -\mathbf{B}(\theta,\rho,I) \| \cr
		 \textrm{subject to:}  \cr
		  f_j(\theta,\rho,I) \leq 0, \quad j= 1,\dots,p\cr
		  g_k(\theta,\rho,I) = 0, \quad k= 1,\dots,q} 
	\end{equation}
where $f_j(\theta,\rho,I)$ are a set of $p$ inequality constraints (e.g. inequality equations defining the boundaries of the geometrical configuration of the current lines), and $g_k(\theta,\rho,I)$ are a set of $q$ equality constraints (e.g. equations to cancel specific $C_{n,total}$ coefficients of the harmonic expansion).

\subsection{Asymmetric cross section}

In the first place, we will search for solutions of the asymmetric configuration. For this purpose, we will consider a dipole, and we will explore a subset of the asymmetric UL magnets based on the following constraints:
\begin{itemize}
	\item All current lines are perpendicular to the Cartesian $xy$-plane along the straight section of the magnet.
	\item The magnitude of the current is the same for all conductors, where only its sign changes. The current can therefore be expressed as 
	\begin{equation*}
		\label{deqn_current1}
		I_j=  s_j I \quad\textrm{with}\quad s_j \in \{-1,1\}
		\
	\end{equation*}
	
	In this regard, the current sign, $s_j$, is chosen to create a positive $B_y$, and therefore $s_j$ is negative on the \textit{right} (i.e. $x>0$) and positive on the \textit{left} (i.e. $x<0$).
	\item All current lines are equidistant to the origin $z=0$, at a distance $\rho$, i.e. radius of the magnet aperture.
	\item The multipole expansion will be considered within a circular domain $D$ centered at $z=0$, and with a radius $R_{ref}$, related to the magnet aperture $\rho$ as $R_{ref} = 2/3 \rho$.
	\item An odd number of current lines, $m_r$, are considered on the \textit{right }(i.e. $x>0$), while an even number of current lines, $m_r+1$, is considered on the \textit{left }(i.e. $x<0$), with a total of current lines $m=2m_r+1$.
	\item A \textit{top-bottom} symmetry is considered, therefore cancelling all skew components of the harmonic coefficients.
	\item From $C_{2}$ all harmonic coefficients must be $0$ up to $C_{n_0}$:
	\begin{equation*}
		\label{deqn_current2}
		C_{n,total}= 0  \quad\textrm{with}\quad n \in \{2,3,\dots,n_0\}
		\
	\end{equation*}
	
	\item The minimum angular distance between adjacent current lines is $\theta_{min}$.
	
\end{itemize}

From the general multipole expansion (\ref{sec:harmonics})

\begin{equation}
	\label{deqn_general}
	\mathbf{B}(z) =  \sum_{n=1}^{\infty} C_{n,total} \left(\frac{z}{R_{ref}}\right)^{n-1}  
	\
\end{equation}
and the expression of the coefficients considering the previous constraints
\begin{equation}
	\label{deqn_coef_total}
	C_{n,total} =  -\frac{\mu_0 I}{2 \pi } \frac{ R_{ref}^{n-1}}{\rho^n} \sum_{j=1}^{m} s_j e^{-i n \theta_j}
	\
\end{equation}
one can see that the global optimization can be expressed in terms of the argument and sign of the current lines
\begin{equation*}
	\label{deqn_minimization1}
	\sum_{j=1}^{m} s_j e^{-i n \theta_j}
	\
\end{equation*}

First, let's find the expression to minimize, which would lead to the maximization of the required $C_{1,total}$ and therefore $\mathbf{B}$. In this case, given the symmetry conditions previously defined, the number of components of $\theta$ is significantly reduced. For an odd number of current lines on the \textit{right}, the only form to maintain a \textit{top-bottom} symmetry is to place a conductor in the mid-plane (i.e. $y=0$). Therefore, for a total of $m$ current lines to form the magnet, where $m=2m_r+1$, the total number of argument variables, $\theta_i$, to characterize the full position of all conductors is $m_r$, for a given $\rho$.

Let's then define an array $\theta$ of size $m_r$ where the argument of the conductors will be stored
\begin{equation*}
	\label{deqn_theta_array}
	\theta = \left\{  \theta_1,\theta_2,\dots \theta_{m_r} \right\}
\end{equation*}
where the first $\left(m_r-1\right)/2$ components will be designating the current lines on the \textit{right}, while the rest will be designating the position of the current lines on the \textit{left }side. Following these indications, and for simplicity of notation, we will designate a number $m_{opt} = \left(m_r-1\right)/2$. We can now write an example of a  general equation $f_{min}$ to be minimized
\begin{equation}
	\label{deqn_fmin}
	f_{min} = -1 + 2 \left( -\sum_{j=1}^{m_{opt}} \cos{\theta_j} + \sum_{j=m_{opt}+1}^{m_r} \cos{\theta_j}  \right)
\end{equation}

In an analogous form, one can find the set of equality constraints to cancel specific harmonic coefficients. Furthermore, with regard to the geometrical constraints, different forms could be established depending of the specific objectives relating to field, conductor limitations, cost, manufacturability, etc. In this example, a minimum angular spacing between conductors was included as
\begin{equation}
	\label{deqn_spacing}
	\min \left\{  \| \theta_j - \theta_{j+1}   \|    \right\} - \theta_{min} \geq 0, \quad j = 1,\dots,m_r-1 \\
\end{equation}

The overall optimization equations can therefore be written as

	\begin{equation}
	\label{deqn_opti_asy}
	\eqalign{\textrm{minimize}  \left( -\sum_{j=1}^{m_{opt}} \cos{\theta_j} + \sum_{j=m_{opt}+1}^{m_r} \cos{\theta_j} \right) \cr
		\textrm{subject to:}  \cr
		\min \left\{  \| \theta_j - \theta_{j+1}   \|    \right\} - \theta_{min} \geq 0, \quad j= 1,\dots,m_r-1   \cr
		 \theta_{min} \leq \theta_j  \leq \frac{\pi}{2}, \quad j = 1, \dots,m_{opt} \cr
		 \frac{\pi}{2}\leq \theta_j  \leq \pi, \quad j = m_{opt}+1, \dots,m_r \cr
		 -1 + 2 \left( -\sum_{j=1}^{m_{opt}} \cos{n\theta_j} + \sum_{j=m_{opt}+1}^{m_r} \cos{n\theta_j}  \right) = 0, \cr 
		 \quad  \quad \quad  \quad  \quad \quad \textrm{with} \quad n = 2, \dots,n_0  } 
	\end{equation}

\begin{figure*}[ht!]
	\centering
	\includegraphics[width=6.5in]{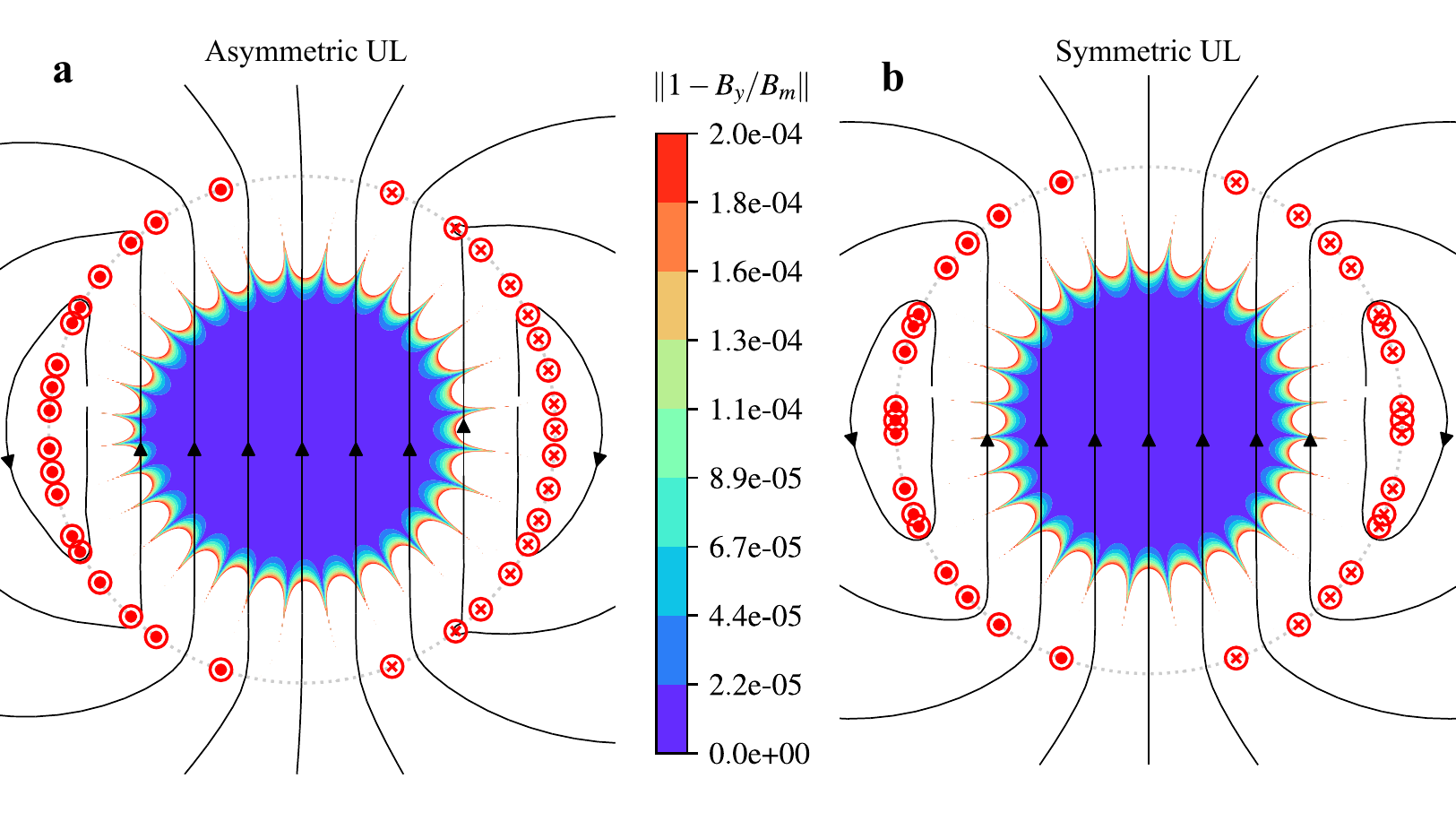}
	\caption{Possible solutions for asymmetric and symmetric configurations with $m_r=17$, $n_0 = 14$ and $\theta_{min}=3^{\circ}$. All the normalized harmonic coefficients $\|c_n\|\leq 5 \cdot 10^{-8}$ units for $n = 2,\dots,14$ for both numerical solutions: (a) Asymmetric configuration. (b) Symmetric configuration, \ref{sec:symmetric}.}
	\label{fig_UL}
\end{figure*}

This problem can be solved numerically taking into consideration the constraints and bounds. An example of a solution to (\ref{deqn_opti_asy}) is illustrated in Figure~\ref{fig_UL}(a). Analogous equations can be formulated to create higher-order asymmetric magnets, e.g. quadrupoles, sextupoles, etc. The results in Figure~\ref{fig_UL}(a) demonstrate that there are solutions to create an asymmetric UL magnet with excellent harmonics, where the normalized harmonic coefficients $\|c_n\|\leq 5 \cdot 10^{-8}$ units for $n = 2,\dots,14$.

\subsection{Surfaces and spatial winding configuration of a single layer}

\begin{figure}[!h]
	\centering
	\includegraphics[width=3.275in]{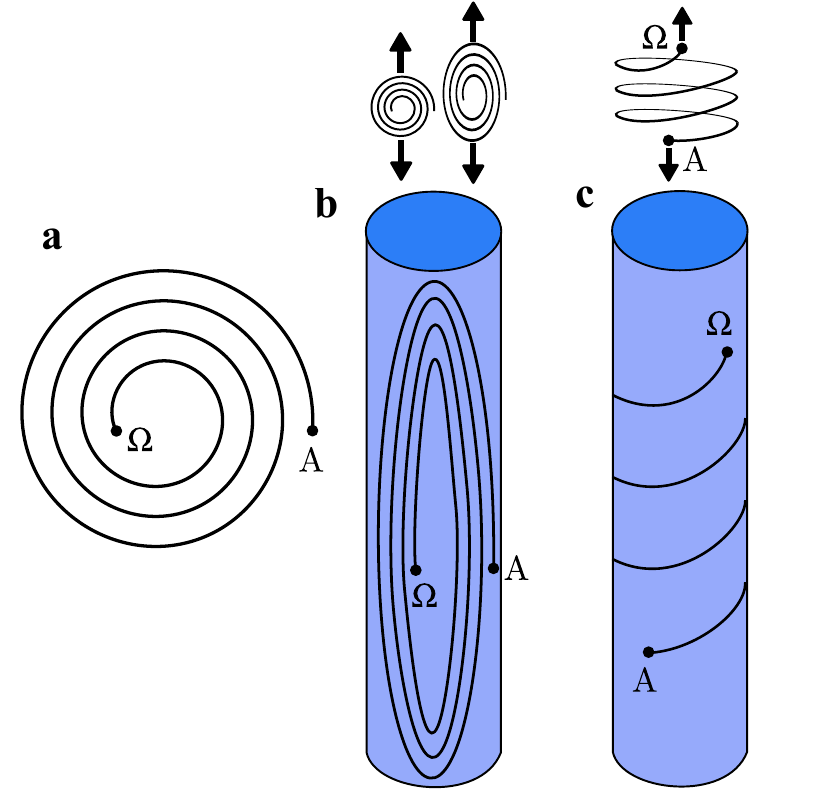}
	\caption{Basic transformations of a spiral geometry to create various winding configurations. (a) Basic spiral curve (b) By stretching and wrapping one could create the CT winding. If we wrap the space curve around a surface with zero Gaussian curvature (e.g. cylindrical surface), once $\Omega$ is located at an \textit{internal} position, the only way out without the conductor crossing itself (e.g. layer jump) is by moving away from the surface where the main curve lies. (c) Transformation into a helix can lead to a solenoid or, by further transformations, to a CCT magnet.}
	\label{fig_winding_basic}
\end{figure}

In the case of a real magnet, the current lines forming the straight section have to be \textit{linked} together, i.e. a continuous conductor is normally used to \textit{wind} the magnet. In most of the magnet configurations (e.g. racetrack coils, CT, block-type coils, solenoids, and CCT) the space curves describing the path of the conductor are based on relatively simple transformations of a spiral geometry that is then wrapped around a certain surface, as illustrated in Figure~\ref{fig_winding_basic}.

In the case of traditional racetrack coil layers (laying on a plane) or CT coil layers (laying on surfaces with zero Gaussian curvature), it is required to depart from this base surface in order to access the beginning and end of the curve simultaneously. It is for this reason that the majority of these types of coils are created with two layers, where an internal layer jump (i.e. change of winding surface) is required.

\begin{figure*}[!h]
	\centering
	\includegraphics[width=6.5in]{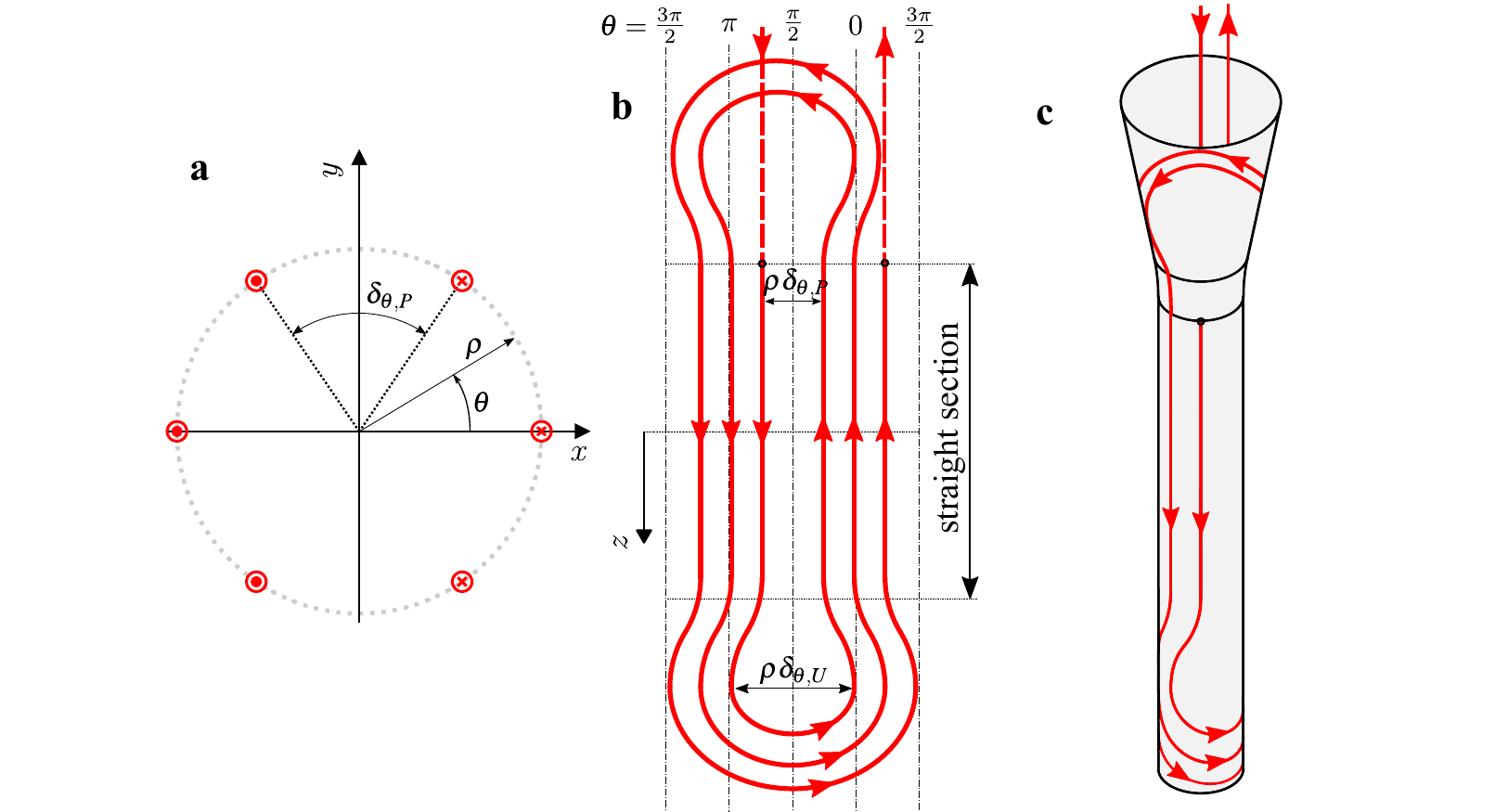}
	\caption{Schematic example of a winding configuration for a symmetric uni-layer dipole magnet. (a) Polar location of the conductor on the straight section of the magnet. (b) Winding configuration in its developed surface. (c) Winding configuration in space lying onto a surface with zero Gaussian curvature.}
	\label{fig_winding_sym}
\end{figure*}

If we try to create a UL symmetric winding configuration, we can see that there are solutions for a space curve $\mathbf{\gamma_{sym}}$ that lies in a surface of zero Gaussian curvature along the straight section, but then it is necessary to transition on the lead end through a negative Gaussian curvature so the conductor can exit the surface. A schematic of a $\mathbf{\gamma_{sym}}$ solution is illustrated in Figure~\ref{fig_winding_sym}, where the conductor \textit{enters} and \textit{exits} the winding surface at the pole region.

On the other hand, now considering asymmetric UL, if one departs from the simple transformations illustrated in Figure~\ref{fig_winding_basic} and considers more complex options, one can find that there are solutions for space curves $\mathbf{\gamma_{asy}}$, describing the path of a single continuous conductor in $\mathbb{R}^3$, that could lie in surfaces of zero Gaussian curvature (e.g. cylindrical surfaces such as cylindrical shells or elliptical cylinders, or most of the surfaces generated by \textit{extruding} closed \textit{two-dimensional} profiles), \textit{entering} and \textit{exiting} the surface through its boundaries with $\mathbf{\gamma_{asy}}$ never \textit{crossing} itself.

\begin{figure*}[!h]
	\centering
	\includegraphics[width=6.5in]{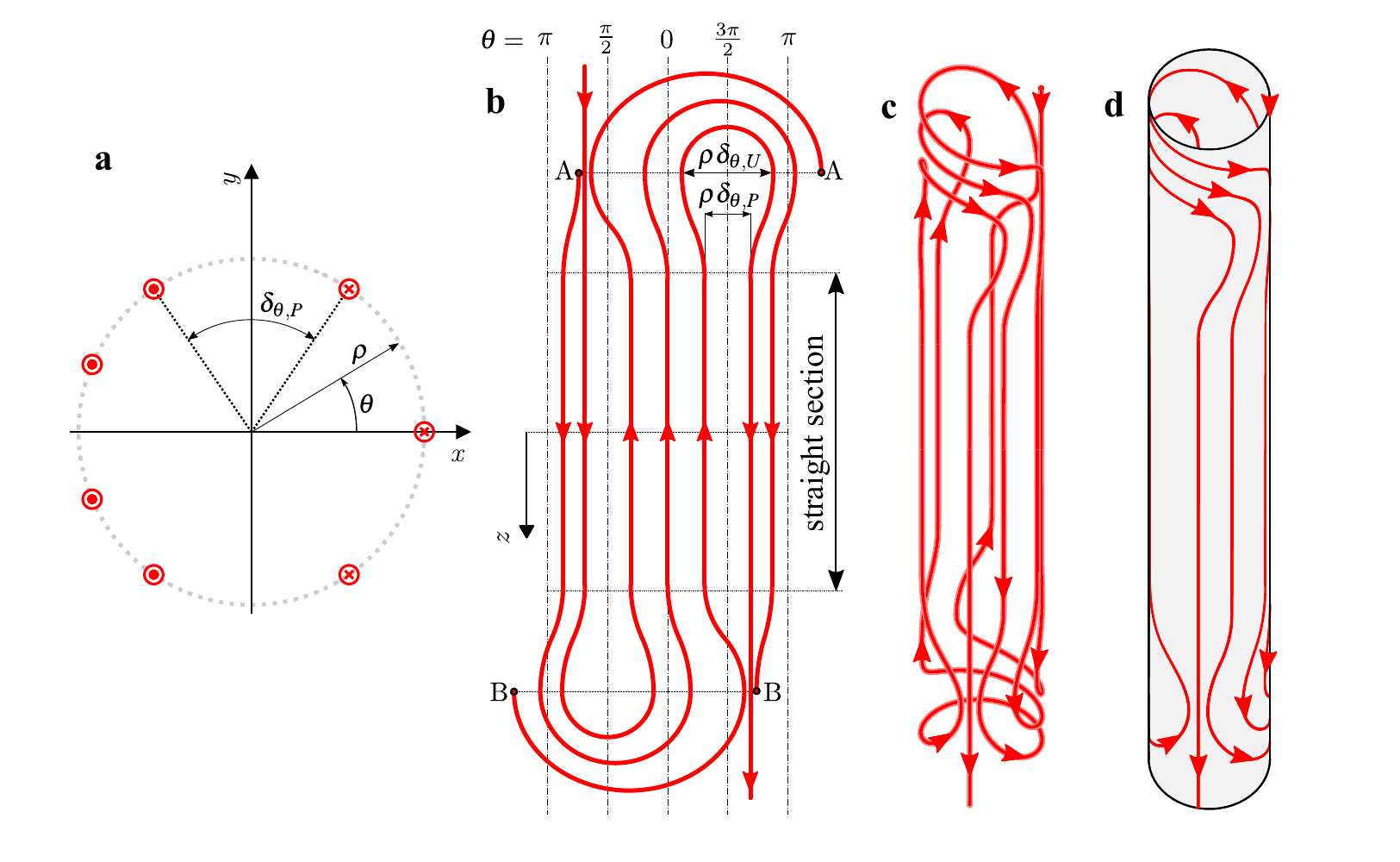}
	\caption{Schematic example of a winding configuration for an asymmetric uni-layer dipole magnet. (a) Polar location of the conductor on the straight section of the magnet. (b)  Winding configuration in its developed surface. In a traditional CT, on each head, there are two pole \textit{tight} turns where a $\delta_{\theta,P}$ is required at $\theta=$ $\pi/2$ and $3\pi/2$. In a UL configuration, only one \textit{tight} turn is required, and therefore the angular space can be expanded from $\delta_{\theta,P}$ to $\delta_{\theta,U}$, where $\delta_{\theta,U} \ge 2\delta_{\theta,P}$. (c)  Winding configuration in space. (d)  Winding configuration in space lying onto a surface with zero Gaussian curvature.}
	\label{fig_winding_main}
\end{figure*}

A schematic example of a solution of an asymmetric UL magnet and its space curve $\mathbf{\gamma_{asy}}$ is illustrated in Figure~\ref{fig_winding_main}. As one can see from this schematic solution, some of the advantages of UL magnets are:
\begin{itemize}
	\item  Reduction by a factor 2 of the number of tight bends described by the conductor around the poles in relation to CT magnets.  
	\item The angular transition $\delta_{\theta}$, which determines the minimum radius of curvature, is \textit{at least} doubled. If we consider $\delta_{\theta,P}$ as the angular space for a CT or SMCT, and $\delta_{\theta,U}$ the angular distance in a uni-layer magnet, we can see that a winding with $\delta_{\theta,U} \ge 2\delta_{\theta,P}$ is possible, Figure~\ref{fig_winding_main}(b). 

\end{itemize}

\begin{figure*}[!h]
	\centering
	\includegraphics[width=6.5in]{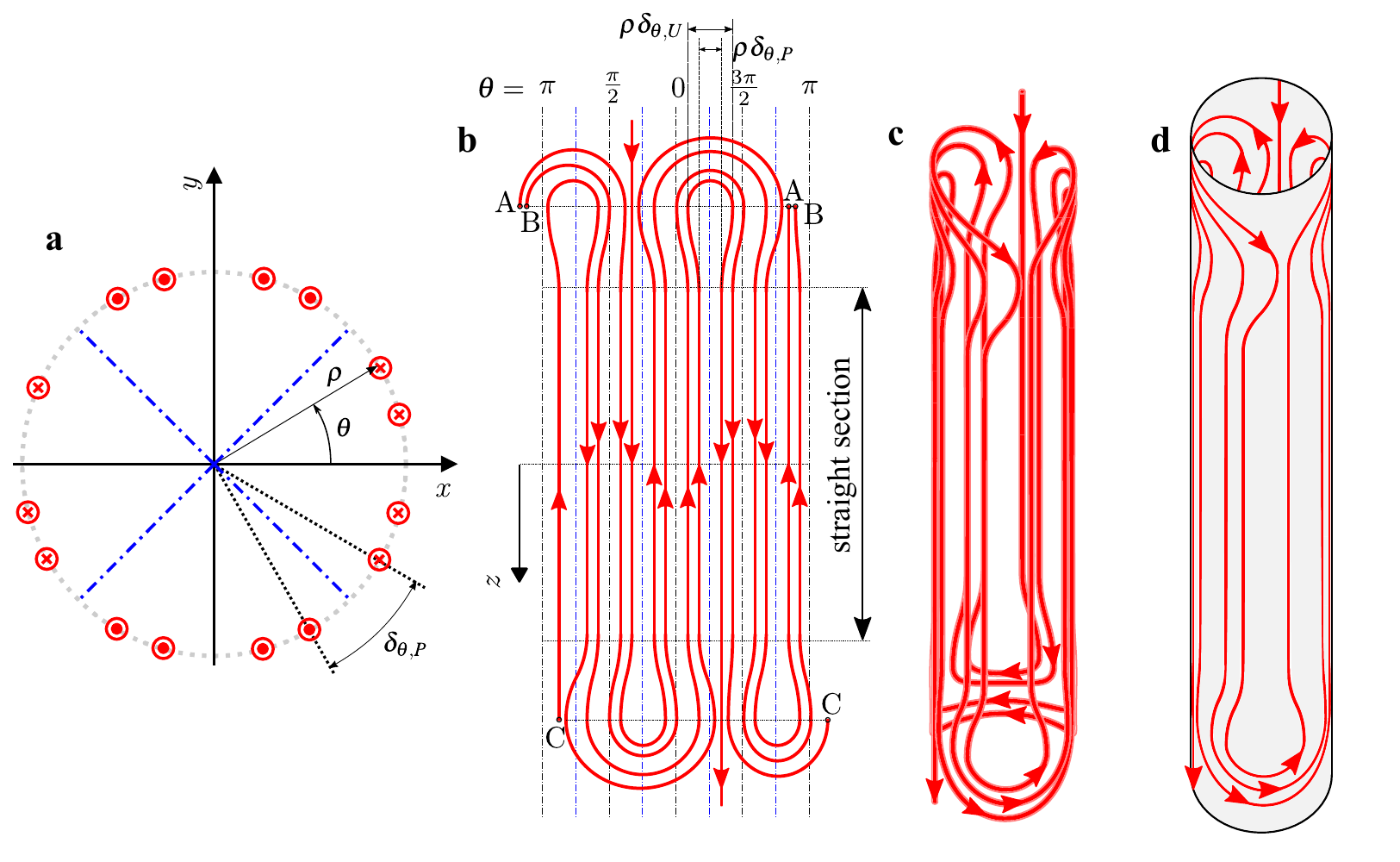}
	\caption{Schematic example of a winding configuration for an asymmetric uni-layer quadrupole magnet. (a) Polar location of the conductor on the straight section of the magnet. (b)  Winding configuration in its developed surface. (c) Winding configuration in space. (d)  Winding configuration in space lying onto a surface with zero Gaussian curvature.}
	\label{fig_winding_quadrupole}
\end{figure*}

It should be noticed that in the UL configuration, one can create a specific magnetic field with a single continuous conductor. This is not only true for the dipole configuration but also for higher order fields, since there are also solutions for  space curves $\mathbf{\gamma_{asy}}$ that satisfy this premise, such as the example illustrated in Figure~\ref{fig_winding_quadrupole} for a quadrupole magnet. Moreover, the winding configurations are also compatible with other surfaces with non-circular apertures.

\subsection{Geometry of the magnet's ends and conductor's bending radius}

In the case of the mathematical representation of the space curves $\mathbf{\gamma_{asy}}$ that satisfy the required conditions schematically illustrated in the previous section, and that are formed by a set of curves that can be defined with parametric continuity $C^2$ (i.e. with continuous zeroth, first and second derivative), one can find solutions using the generalized superelliptical curves wrapped around $\mathbb{R}^3$ surfaces with zero Gaussian curvature. If one considers a cylindrical surface, with its axis oriented along the \textit{z-axis} of a Cartesian coordinate system, the expression of a wrapped generalized superellipse can be written as
\begin{equation}
	\label{deqn_gsellipse}
	\eqalign{
	\mathbf{\gamma_{gse}}(\Phi)= &\rho \cos{\left(\theta_h - \delta_i \sin^{\frac{2}{\chi}}\left(\Phi \right)\right)}\hat{\mathbf{i}}  \\ 
	&+ \rho \sin{\left(\theta_h - \delta_i \sin^{\frac{2}{\chi}}\left(\Phi \right)\right)}\hat{\mathbf{j}}\\
	&+ \left(z_0 - l_t \cos^{\frac{2}{\zeta}}\left(\Phi \right)\right)\hat{\mathbf{k}}  \quad \textrm{with} \; \Phi \in \left(0,\pi/2 \right)\\}
\end{equation}
where $\rho$ is the radius of the cylinder where $\mathbf{\gamma_{gse}}$ lies, $\zeta$ and $\chi$ are the parameters defining the order of the superellipse, $\theta_h$ and $z_0$ position the curve within the surface, $\delta_i$ and $l_t$ define the size of the curve, and $\Phi$ is the parametrization variable.

We can also consider a simplified case of the generalized form, in which $\zeta = \chi$ and where the center of the superellipse is at $x=0$ within the Cartesian coordinate system. In this case the mathematical expression can be simplified as

\begin{equation}
	\label{deqn_sellipse}
	\eqalign{
	\mathbf{\gamma_{se}}(\varphi)= &\rho \sin{\left(\frac{a \cos^{\frac{2}{\zeta}}{\left(\varphi \right)}}{\rho} \right)}\hat{\mathbf{i}}\\
	&+ \rho \cos{\left(\frac{a \cos^{\frac{2}{\zeta}}{\left(\varphi \right)}}{\rho} \right)}\hat{\mathbf{j}} \\
	& + b \sin^{\frac{2}{\zeta}}{\left(\varphi \right)}\hat{\mathbf{k}} \quad\textrm{with}\quad \varphi \in \left(0,\pi/2 \right)\\
}
\end{equation}
where
\begin{equation*}
	\label{deqn_one}
	a = \rho\frac{\delta_{\theta}}{2}\\
\end{equation*}

Similar curves to  $\mathbf{\gamma_{se}}$ are also used as base in traditional CT coils to minimize the strain energy of the winding block in the head of the coils \cite{Auchmann2004}.

The curves $\mathbf{\gamma_{gse}}$ and $\mathbf{\gamma_{se}}$ can be used to describe the curvature of the conductor in the UL's head as schematically represented in Figure~\ref{fig_winding_main}. Furthermore, as it was described in the previous section, the spatial configuration of UL magnets allows for a significant increase of $\delta_{\theta}$, which ultimately leads to a higher radius of curvature induced in the conductor.

If we consider a set of $\mathbf{\gamma_{se}}$ curves to describe the magnet's end, we can find the optimal parameters of the curve (i.e. $b$ and $\zeta$) that will maximize the radius of curvature in the conductor (\ref{sec:magnet_end_geometry}). In this regard, when optimizing the conductor's curvature at the magnet's end, algorithms normally rely on the minimization of the strain energy \cite{Auchmann2004} of the conductor.  In the case here explored, as we are considering a simplification of an idealized conductor, we will follow a strategy to numerically determine the optimal value of variables defining the space curves that maximize the minimum radius of curvature along the end of the magnet, $R_{min}$, for a given $\rho$ and $\delta_{\theta}$. Considering the previous parameters and $\varphi$ as the parametrization variable of $\mathbf{\gamma_{se}}$ (\ref{sec:magnet_end_geometry}), we can write the optimization problem to determine the optimal parameters as

\begin{equation}
	\label{deqn_max}
	\max_{b\in \mathbb{R} \mid b>0, \zeta\in \mathbb{R} \mid \zeta \geq 2} \quad \min_{\varphi\in \left(0,\pi/2 \right)}\ \{R_{min}\} \\
\end{equation}

\begin{figure}[t]
	\centering
	\includegraphics[width=3.275in]{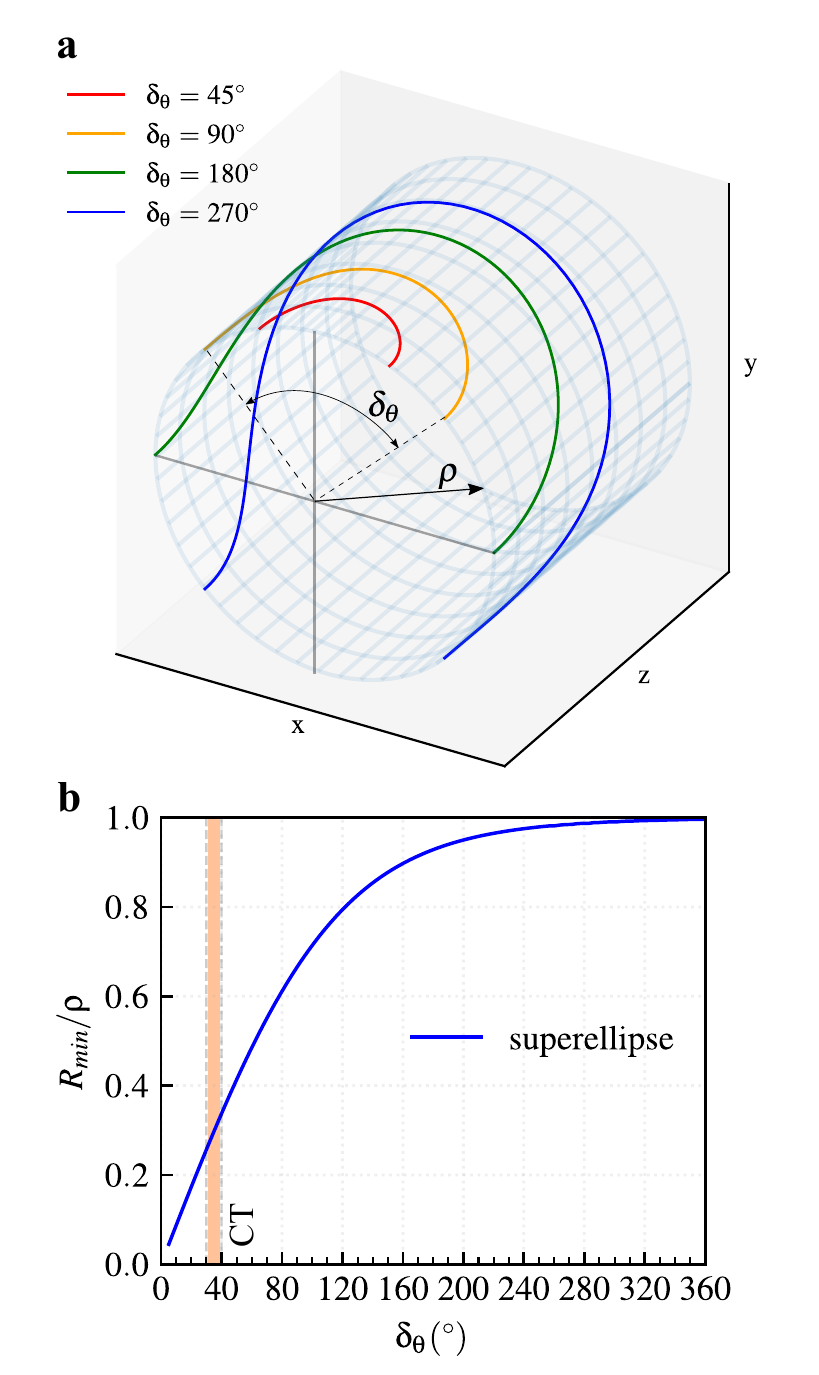}
	\caption{ Superelliptical curves wrapped around a cylindrical surface in the pole region. (a) Magnet end section for example cases of various $\delta_{\theta}$. (b) Ratio between the highest minimum radius of curvature $R_{min}$ and the radius of curvature of the supporting surface $\rho$ that is possible during the end transition between two straight section lines, as a function of  $\delta_{\theta}$. The typical range of $\delta_{\theta}$ for traditional CT magnets is $30$ to $40^{\circ}$ \cite{Todesco2013}.}
	\label{fig_end}
\end{figure}

Figure~\ref{fig_end}a illustrates the superelliptical optimal curves $\mathbf{\gamma_{se}}$ for different example cases of $\delta_{\theta}$, i.e. curves with the maximum radius of curvature that can be achieved by a superellipse for a given set of $\rho$ and $\delta_{\theta}$. The optimal solution, as previously described, can be expressed as the dimensionless parameter $R_{min}/\rho$ as a function of the available $\delta_{\theta}$, as illustrated in Figure~\ref{fig_end}(b). This dimensionless parameter relates the ratio between the radius of the cylinder (i.e. radius of the aperture) and the minimum radius of curvature required in the conductor along the \textit{pole turn}, for a given value of $\delta_{\theta}$. In this regard, the value of $\delta_{\theta}$ in the innermost layer of traditional dipole CT magnets is between $30$ to $40^{\circ}$ \cite{Todesco2013} (about $30^{\circ}$ for both the LHC main dipole \cite{Rossi2004}, and the new D1 beam separation dipole \cite{Xu2013}). 

If we consider that a higher $\delta_{\theta}$ is possible in UL magnets (i.e. $\delta_{\theta,U} \ge 2\delta_{\theta,P}$, Figure~\ref{fig_winding_sym} and Figure~\ref{fig_winding_main}) than in traditional CT magnets, from Figure~\ref{fig_end}b we could see that a higher value of the ratio could also be achieved. Moreover, if we consider that a conductor has a minimum radius of curvature $R_{min,c}$, after which we start degrading it, we can see that a higher ratio translates into the possibility of winding a much smaller coil in a UL configuration in relation to what it would be possible in a CT coil.

In the UL configuration, in order to maximize the radius of curvature within the conductor, it is necessary to maximize the available $\delta_{\theta}$ for the most critical \textit{turn}. For this purpose, we could create a set of \textit{transitions} from the straight section to the end-section described by a set of $\mathbf{\gamma_{se}}$ curves. This is possible through a set of $\mathbf{\gamma_{gse}}$ curves. Therefore, the complete geometry of the space curves $\mathbf{\gamma_{asy}}$ (i.e. asymmetric UL), can be described by joining together three types of curves:
\begin{enumerate}
	\item A straight line along the straight section of the UL.
	\item A superelliptical curve $\mathbf{\gamma_{se}}$ describing the turn around the pole region, i.e. end section.
	\item A generalized superelliptical curve $\mathbf{\gamma_{gse}}$ describing the transition between the straight section to the pole curve, i.e. transition section.
\end{enumerate}

\begin{figure*}[!h]
	\centering
	\includegraphics[width=6.5in]{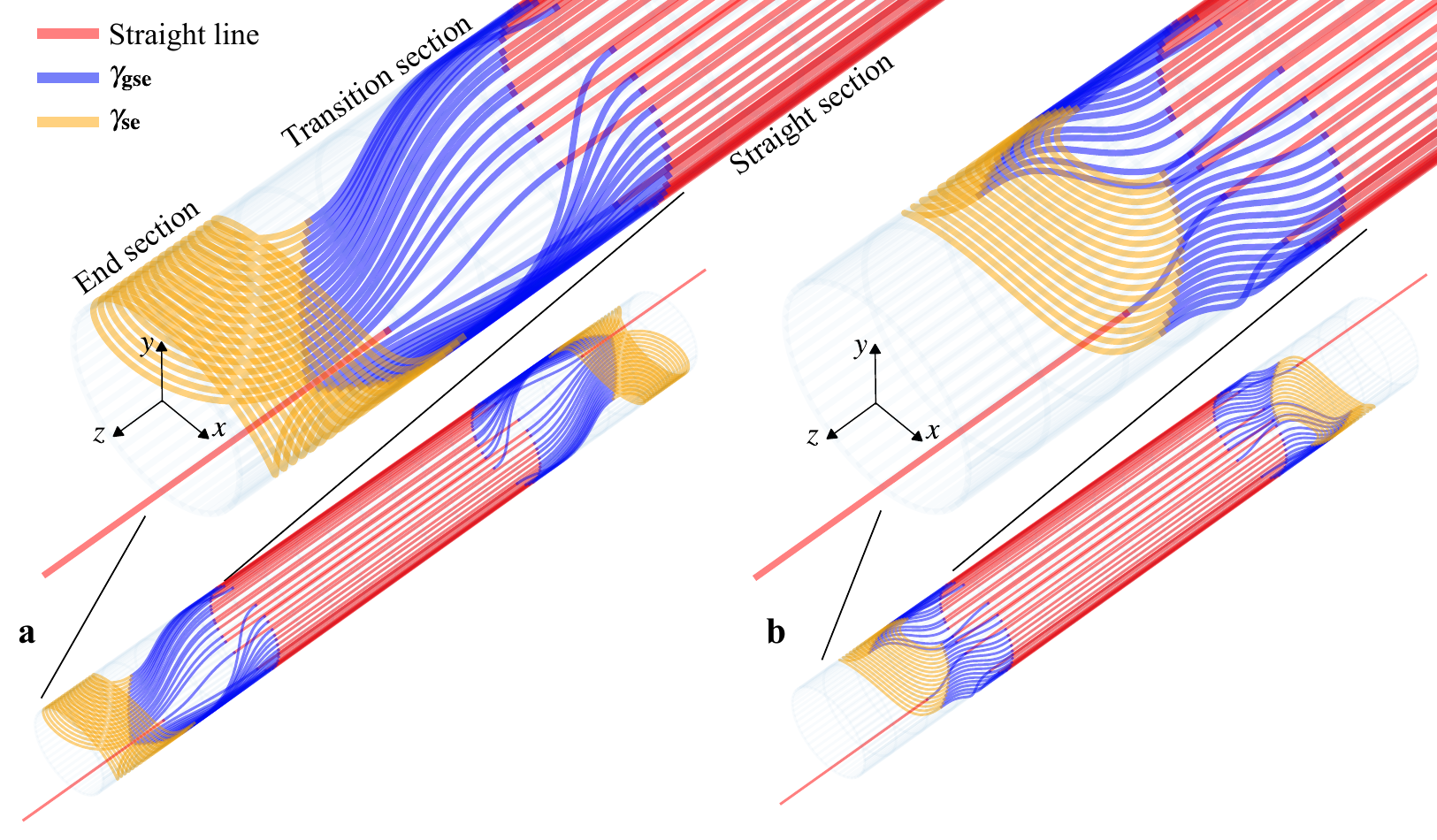}
	\caption{ Two potential 3D solutions (i.e. sets of straight, $\mathbf{\gamma_{gs}}$ and $\mathbf{\gamma_{gse}}$ curves) to the asymmetric UL illustrated in Figure~\ref{fig_UL}a, with $m_r=17$, $n_0 = 14$ and $\theta_{min}=3^{\circ}$. The minimum $\theta_{min}=3^{\circ}$ is conserved at the straight section and beginning and end of the transition section. (a) Solution representing the optimized curves that maximize the radius of curvature of the conductor. (b) Another example of a solution for the same \textit{straight section} configuration, in which a lower radius of curvature in the conductor is permitted.}
	\label{fig_curves_link}
\end{figure*}

A detailed view of an example of how the curves are linked together is illustrated in Figure~\ref{fig_curves_link}. Depending on the optimization objectives (e.g. field and field quality in the ends, minimization of curvature, or minimization of layer overall length), the end's curvature can be design accordingly.

\subsection{Multi-layer \textit{uni-layer} magnets}
\begin{figure}[t]
	\centering
	\includegraphics[width=3.275in]{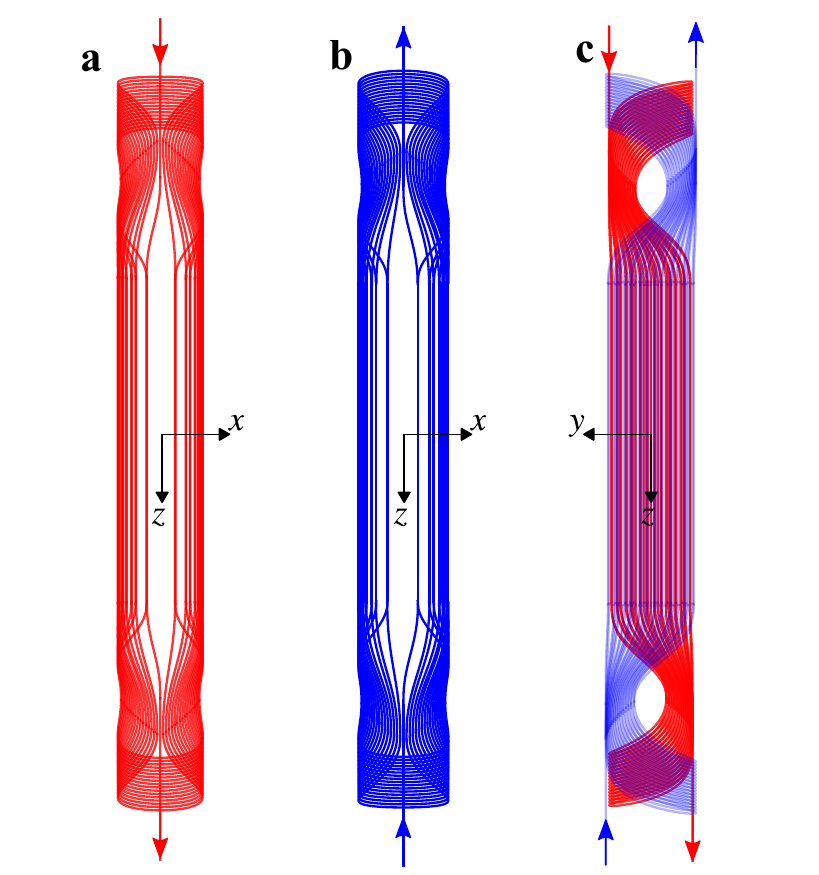}
	\caption{Multi-layer UL magnet. (a) Layer~1 with an odd number of current lines in the region $x>0$. (b) Layer~2 with an odd number of current lines in the region $x<0$. (c) Assembled Layer~1 and 2.}
	\label{fig_two_layer}
\end{figure}

Creating \textit{multi-layer} UL magnets is also possible. In this regard, one could optimize the conductor's position independently for each layer or as a whole assembly. One could also take advantage of alternating the position of the odd number of conductors from \textit{left} to \textit{right}, also optimizing the field at the ends of the magnet, Figure~\ref{fig_two_layer}. UL magnets also allow for the creation of combined function magnets by assembling multiple layers of various types, e.g. a dipole layer 1 within a quadrupole layer 2.

\section{Comparison of magnet concepts}

In this section, we will try to establish a fair comparison of the various dipole magnet concepts based on how efficiently an ideal conductor can be used to form a magnetic field, the minimum radius of curvature and the ability to create coils for very high field magnets. We will also investigate the main advantages of the UL concept in terms of developing and fabricating magnets in relation to the other mentioned concepts.

First, if we look at the magnet design process, when aiming at a particular aperture $\rho$, we can maximize the efficiency of the use of conductor by placing it as close as possible to the domain in which we plan to create the field. In this regard, and considering the geometry of the CT, CCT, and UL configurations, there are limitations on our potential design, since there is a limit on the minimum radius of curvature $R_{min,c}$ at which we can bend a conductor before we degrade it. Consequently, only designs where the \textit{winding} minimum radius of curvature ($R_{min}$) is higher than the conductor minimum radius ($R_{min,c}$) are possible. This is especially relevant for HTS conductors such as {\sc corc}\textsuperscript{\textregistered}.

In the CT concept, the minimum radius that is achieved while winding is related to the radius of the aperture $\rho$ and the angular space in the innermost turns of the pole region $\delta_{\theta,P}$. In the case of the UL concept, the same is true, but as we have shown before, a significantly higher value of $\delta_{\theta,U} \ge 2\delta_{\theta,P}$ can be achieved, Figure~\ref{fig_winding_main}(b). If we examined the solution illustrated Figure~\ref{fig_curves_link}(a) we can see this difference is even greater. If we take as a reference for the CT magnet a value of $\theta \approx 70^{\circ}$, leading to a value of $\delta_{\theta,P} \approx 40^{\circ}$  ($2\left(90-70)\right)=40$), for the solution illustrated in Figure~\ref{fig_curves_link}a we have $\delta_{\theta,U} \approx 5.6\delta_{\theta,P}$.

In the case of the CCT (\ref{sec:CCT}), it is possible to achieve a relatively high $R_{min}/ \rho$ ratio by increasing the inclination of the conductor $\alpha$. However, by increasing $\alpha$, the value of the dipole field $B_{y,CCT}$ rapidly decreases.

\begin{figure}[!ht]
	\centering
	\includegraphics[width=3.275in]{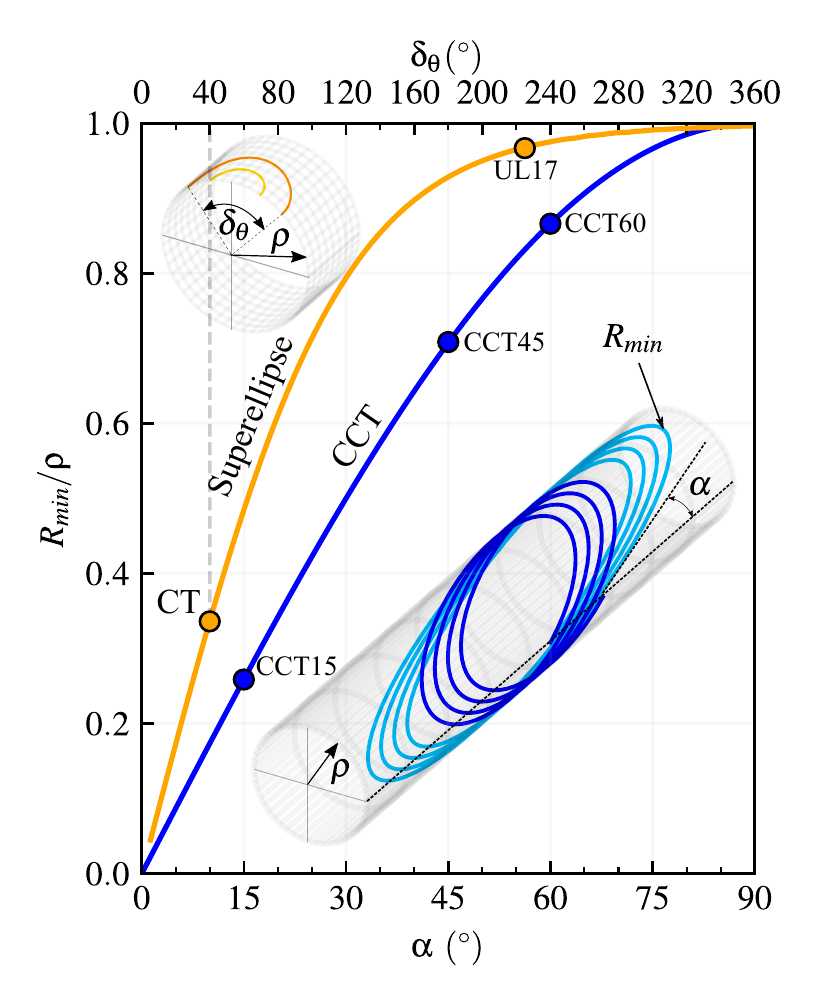}
	\caption{Ratio between the minimum radius of curvature along the curve, $R_{min}$, and the radius of curvature of the cylinder, $\rho$, for the superellipse as a function of $\delta_{\theta}$, and for a canted cosine theta as a function of the inclination $\alpha$. The points represent some representative cases: CT represents a typical $cos\theta$ magnet with $\delta_{\theta}=40^{\circ}$, CCT15, CCT45 and CCT60 represent CCT magnets with a corresponding inclination of $\alpha = 15, 45 \: \textrm{and} \: 60^{\circ}$ respectively, while UL17 represents the 3D solution illustrated in Figure~\ref{fig_curves_link}a, with $m_r=17$, $n_0 = 14$ and $\theta_{min}=3^{\circ}$.}
	\label{fig_angle_comparison}
\end{figure}

The ratio $R_{min}/\rho$ as a function of the main driving parameters for the discussed magnet concepts is illustrated in Figure~\ref{fig_angle_comparison}.

It can be seen how the ratio for CT is about 0.34 (for most of UL designs the ratio should be considerably higher), and considering a CCT with $\alpha = 15^{\circ}$ (the optimal $\alpha$ is normally considered between 10 to $20^{\circ}$ \cite{Brouwer2015}) we can see that the ratio is about 0.26. Assuming the same aperture and conductor, these ratios indicate that it will be easier to wind a UL coil than a CT coil, and significantly easier than a CCT coil since the required curvature of the conductor is minimized in the UL concept. If we consider the example illustrated in Figure~\ref{fig_curves_link}a, the value of the ratio $R_{min}/ \rho$ was approximately 0.97. If we are to design a magnet with a conductor with an intrinsic minimum radius of curvature $R_{min,c}$, the minimum aperture of the coil that we can create is about 2.85 times higher in the case of a CT (or SCMT) and 3.73 higher in the case of a CCT with $\alpha = 15^{\circ}$, all in relation to the illustrated UL.

This geometrical effect is of even more profound importance when examining potential real applications using HTS in very high field magnets, where its use is normally considered within the high field region, i.e. innermost layers. In the case of accelerator magnets, the expectation is to be able to wind coils at an aperture of about 50 mm, i.e. $\rho = 25 \, \textrm{mm}$. This is especially challenging for HTS conductors that easily degrade with a tight radius of curvature, such as REBCO, and derived wires such as {\sc corc}\textsuperscript{\textregistered}. 

If we take the {\sc corc}\textsuperscript{\textregistered} wire as an example, where the reported $R_{min,c}$ is about 20 to 25 mm \cite{Weiss2017,VanDerLaan2019,Wang2021,Stern2022}, we can then compute the smallest aperture that we can wind with the wire without inducing any degradation, for every different magnet concept. The results are summarized in Table~\ref{tab:concept_comparison}.

\begin{table*}[]
	\caption{Smallest possible aperture diameter (i.e. $2\rho$) that can be wound without degrading the conductor assuming a minimum bending radius of $R_{min,c} = 25\, \textrm{mm}$ (i.e. similar to {\sc corc}\textsuperscript{\textregistered} wire). In order to compare the dipole field transfer function between the various coil designs, the same length of ideal line conductor is used to create 1 m of straight section, \ref{sec:comparison_UL}. The dipole field transfer function for the CT design is computed assuming an equivalent cross section to the UL configuration.}
\centering
\begin{tabular}{p{1.6in} p{1in} p{1.75in} p{1.4in}}
	\toprule
	Coil design & Smallest possible aperture (mm) & Dipole field transfer function for the same constant length of conductor across all coil designs (T/kA) & Central field ratio ($B_m/B_{m,UL17}$)\\
	\midrule
		UL (UL17 from Figure~\ref{fig_curves_link}(a))& 51 & 0.218 & 1\\
		CT ($\delta_{\theta} = 40^{\circ}$)  & 147 & 0.075 & 0.34\\
		CCT ($\alpha$ = 15$^{\circ}$)  &  192 & 0.051 & 0.23\\ 
		CCT ($\alpha$ = 45$^{\circ}$) &   71 & 0.081 & 0.37\\ 
		CCT ($\alpha$ = 60$^{\circ}$) &   58 & 0.064 & 0.29\\ 
	\bottomrule
\end{tabular}
\label{tab:concept_comparison}
\end{table*}

\begin{table*}[]
	\caption{Minimum radius of curvature $R_{min}$ required in the conductor to wind a coil with an aperture of 50 mm. The dipole field transfer function for the CT design is computed assuming an equivalent cross section to the UL arrangement.}
	\centering
	\begin{tabular}{p{1.6in} p{1in} p{1.75in} p{1.25in}}
		\toprule
		Coil design & Minimum radius of curvature (mm) &  Dipole field transfer function for the same constant length of conductor across all coil designs (T/kA) & Central field ratio ($B_m/B_{m,UL17}$)\\
		\midrule
		UL (UL17 from Figure~\ref{fig_curves_link}(a))& 24 & 0.222 & 1\\
		CT ($\delta_{\theta} = 40^{\circ}$)  & 8.5 & 0.222 & 1\\
		CCT ($\alpha$ = 15$^{\circ}$)  &  6.5 & 0.197 & 0.89\\ 
		CCT ($\alpha$ = 45$^{\circ}$) &   17 & 0.115 & 0.52\\ 
		CCT ($\alpha$ = 60$^{\circ}$) &   21 & 0.075 & 0.34\\
		\bottomrule
	\end{tabular}
	\label{tab:concept_comparison_curvature}
\end{table*}

We can see that the CT design, and CCT options with low inclination, can only be wound around a much larger aperture in comparison to UL. Moreover, although the CCT design allows for an aperture that is relatively close to what is possible by the UL design, this is achieved at the expense of significantly increasing the inclination angle $\alpha$, which results in a very inefficient design (\ref{sec:comparison}), as it can be seen from the dipole field transfer function. In this regard, the UL design could allow for a significant increase in the field generated in relation to CCT magnets with relatively high $\alpha$ (using the same amount of conductor).

If one considers the inner layer of very high field magnets, the high geometrical constraints discussed above might make UL magnets, not only the optimal option but perhaps one of the only viable solutions to wind a small coil with REBCO-based wire around the \textit{small} aperture. 

The geometrical advantage of UL is also translated into the significant increase in the minimum radius of curvature required to wind a coil for a specific aperture. This is relevant for HTS and LTS conductors, reducing the probability of mechanical instabilities in the cable due to winding \textit{tight} turns. The relative minimum radius of curvature for winding a 50 mm diameter aperture for the various concepts is summarized in Table~\ref{tab:concept_comparison_curvature}, illustrating the significant improvement in the case of UL magnets.


\subsection{Discussion}

The UL concept presents some considerable advantages with respect to other concepts in terms of available design options, magnet development time, simplicity of assembly, and mass production.

\subsubsection{Design}

As previously discussed, the UL magnet can be designed as a sector-type coil (as in a SMCT), or as an individual turn type (as in the CCT concept), or as a combination of both, therefore providing full flexibility to address a particular set of constraints. In particular, the UL magnet allows for creating high order magnetic fields within a single layer (and a single conductor length), also leading to the possibility of creating UL magnets with an odd number of layers. 

 In relation to CT or SMCT coils, the number and degree of tight turns around the pole region in the magnet's ends is significantly reduced, as well as eliminating the hard bend required for the layer-jump in the high field region, which might lead to a reduction of potential issues in these regions. 

\subsubsection{Fabrication}

\begin{figure}[!h]
	\centering
	\includegraphics[width=3.275in]{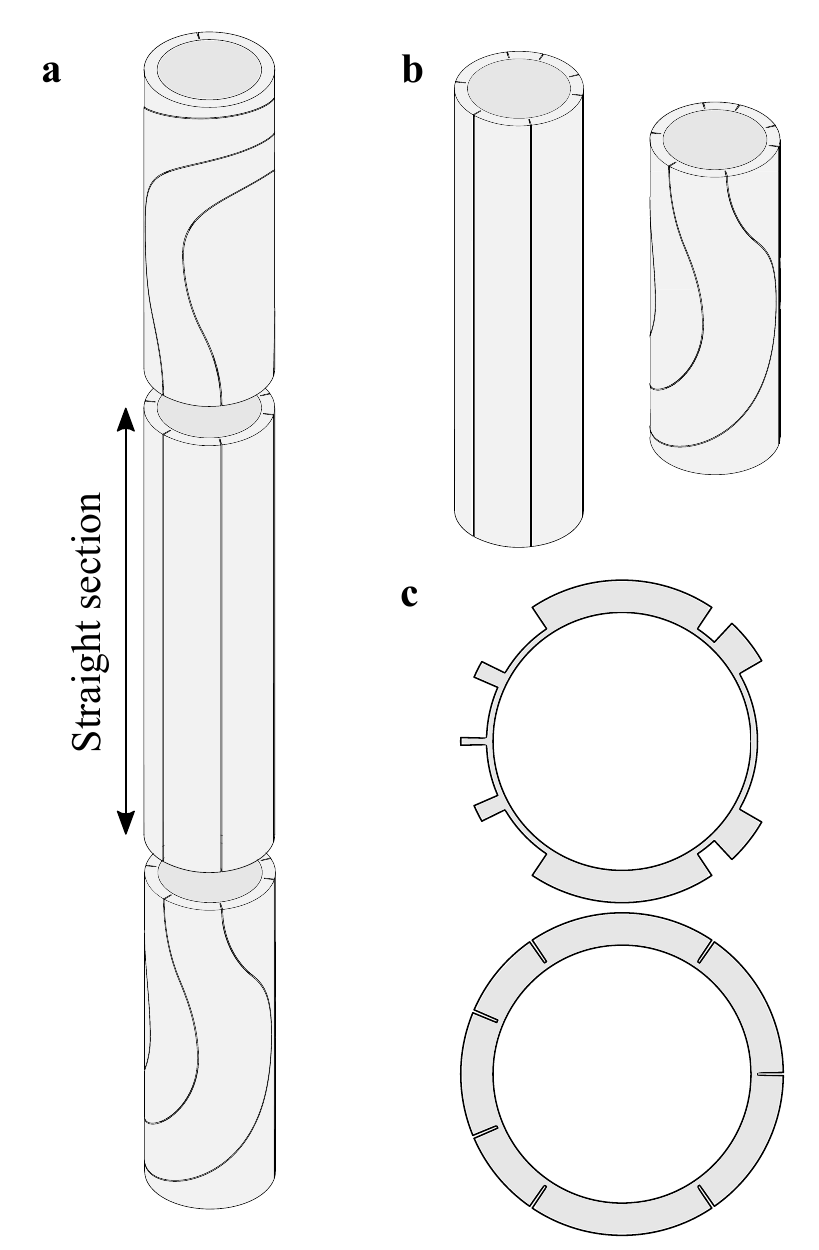}
	\caption{Schematic representation of the UL concept support structure. (a) The winding mandrel can be manufactured in a single piece or split into different sections. (b) If one considers the split between the straight section and the magnet's ends, one could see that the straight section should have a much lower production cost than a CCT magnet. (c) The UL concept can accommodate a large variety of design options, including grooves for individual turns or for sector-type winding.}
	\label{fig_manufacturing}
\end{figure}

The main advantage in relation to other designs is the potential of simplification of the manufacturing of the mandrel. Some of the various options when manufacturing the mandrel are schematically represented in Figure~\ref{fig_manufacturing}. The UL mandrel can be manufactured as a single piece or split between the straight section and the magnet's ends. This option opens a significant reduction of the overall cost of the mandrel, since its straight section is simply an extruded volume. This represents a significant advantage over CCT magnets, where producing long mandrels is a challenge. The type of simple straight structure required for the UL concept could be produced either by longitudinal machining of tubes, metal extrusion, or even as a laminated structure joined by longitudinal elements, leading to an overall reduction of cost, especially with regard to very long magnets.

\section{Conclusion}
The novel concept of \textit{Uni-layer} magnets has been presented. The asymmetric \textit{Uni-layer} magnet shows significant advantages over other concepts, merging together some of the benefits of the CT and CCT, providing a high quality field in terms of harmonics, within a single-layer (eliminating the need of internal layer jumps), using a single continuous conductor, and with a higher minimum radius of curvature required in the conductor during winding. This transformative new concept has the potential to enable and accelerate the adoption of HTS in very high field superconducting magnets for high energy accelerator and other applications.

\ack

 The work at LBNL was supported by the U.S. Department of Energy, Office of Science, Office of High Energy Physics, through the US Magnet Development Program under Contract No. DEAC02-05CH11231.

\newpage
\clearpage
\appendix

\section{Harmonics of the magnetic field}
\label{sec:harmonics}

From Maxwell's equations in differential form 
	\begin{equation*}
	\label{deqn_Maxwell}
	\eqalign{
		\nabla \cdot \mathbf{E} &= \frac {\rho} {\varepsilon_0}   \\
		\nabla \times \mathbf{E} &= -\frac{\partial \mathbf{B}} {\partial t}   \cr
		} 
	\qquad
	\eqalign{
		\nabla \cdot \mathbf{B} &= 0   \\
		\nabla \times \mathbf{B}& = \mu_0\left(\mathbf{J} + \varepsilon_0 \frac{\partial \mathbf{E}} {\partial t} \right)    \cr
	} 
\end{equation*}
and considering a region of space free of charges, the magnetic field $\mathbf{B}$ can be described as 
\begin{equation}
	\label{deqn_simple1}
	\nabla \cdot \mathbf{B} = 0   \quad\textrm{and}\quad  \nabla \times \mathbf{B} = 0
\end{equation}

In addition, if we consider that the magnetic field  $\mathbf{B}$ is within a cylindrical aperture centred at the origin, i.e. $x=y=0$, and that it is created through a series of idealized infinite current lines circulating parallel to the \textit{z-axis} of a Cartesian coordinate system (i.e. perpendicular to the $xy$-plane), we also have
\begin{equation}
	\label{deqn_simple2}
	\frac{\partial B_x} {\partial z} = \frac{\partial B_y} {\partial z} = \frac{\partial B_z} {\partial z} = 0   
\end{equation}

Considering (\ref{deqn_simple1}) and (\ref{deqn_simple2}) one can prove that $B_x(x,y)$ and $B_y(x,y)$ satisfy the Cauchy-Riemann equations
\begin{equation*}
	\label{deqn_simple3}
	\eqalign{
	\frac{\partial B_x} {\partial y} + \frac{\partial B_y} {\partial x}  &= 0 \\
	\frac{\partial B_x} {\partial x} - \frac{\partial B_y} {\partial y}  &= 0 }
\end{equation*}
and therefore $\mathbf{B}$ can be written as a function of complex variable $z$ in the complex plane 
\begin{equation*}
	\label{deqn_simple4}
	\mathbf{B}(z) =  B_y(x,y) + i B_x(x,y)   \quad\textrm{where}\quad z = x + iy
	\
\end{equation*}
that can also be expressed as a power series
\begin{equation*}
	\label{deqn_simple5}
	\mathbf{B}(z) =  \sum_{n=0}^{\infty} C_n z^n    \quad\textrm{with}\quad z \in D
	\
\end{equation*}
over a circular domain D defined with a radius equal to the minimum distance between any current line and the center of the expansion, i.e. $z=0$. The coefficients $C_n$ are known as the multipolar expansion coefficients or harmonic coefficients,
\begin{equation*}
	\label{deqn_simple6}
	C_n = B_n+i A_n 
	\
\end{equation*}
where their real ($B_n$) and imaginary ($A_n$) parts are normally referred to as the normal and skew coefficients respectively. In magnets, the multipole expansion series is usually expressed as
\begin{equation}
	\label{deqn_multipole}
	\mathbf{B}(z) =  \sum_{n=1}^{\infty}C_n \left(\frac{z}{R_{ref}}\right)^{n-1}    \quad\textrm{with}\quad z \in D
	\
\end{equation}
where $R_{ref}$ is the reference radius, normally considered as 2/3 of the aperture radius ($\rho$) of the magnet. The harmonic coefficients can also be normalized as
\begin{equation*}
	\label{deqn_simple6}
	c_n = b_n+i a_n = 10^4 \frac{C_n}{B_m} 
	\
\end{equation*}
where $B_m$ is normally considered as the main field.

\subsection{Magnetic field and harmonics of an infinitely long current line}
Let's imagine an idealized infinitely long current line perpendicular to the $xy$-plane, which is considered as the complex plane with $z=x + iy$. From the Biot-Savart law, the magnetic field $\mathbf{B}$ created at a point $z$ by a current $I$ located at point $z_s$ is given by 
\begin{equation}
	\label{deqn_line}
	\mathbf{B}(z) =  \frac{\mu_0 I}{2 \pi} \frac{1}{z-z_s}
	\
\end{equation}

This equation can be developed into an expression that includes the complex power series
\begin{equation*}
	\label{deqn_line_2}
	\eqalign{
		\mathbf{B}(z) &=  \frac{\mu_0 I}{2 \pi} \frac{1}{z-z_s} = -\frac{\mu_0 I}{2 \pi} \frac{1}{z_s-z} \\
		& = -\frac{\mu_0 I}{2 \pi z_s} \frac{1}{1-\frac{z}{z_s}} = -\frac{\mu_0 I}{2 \pi z_s}\sum_{n=0}^{\infty} \left(\frac{z}{z_s} \right)^{n} \\
		&= -\frac{\mu_0 I}{2 \pi z_s}\sum_{n=1}^{\infty} \left(\frac{z}{z_s} \right)^{n-1}}
\end{equation*}

One can further manipulate the equation in order to reach an equivalent expression to (\ref{deqn_multipole}) 
\begin{equation*}
	\label{deqn_line_2}
	\eqalign{
		\mathbf{B}(z) &=  \sum_{n=1}^{\infty} -\frac{\mu_0 I}{2 \pi z_s} \left(\frac{z}{z_s} \right)^{n-1} \\
		&= \sum_{n=1}^{\infty} -\frac{\mu_0 I}{2 \pi} \frac{1}{z_s^n} \left(\frac{R_{ref}}{R_{ref}} \right)^{n-1} \left(z \right)^{n-1} \\
		&= \sum_{n=1}^{\infty} -\frac{\mu_0 I}{2 \pi} \frac{R_{ref}^{n-1}}{z_s^n} \left(\frac{z}{R_{ref}} \right)^{n-1}}
	\
\end{equation*}
leading to the general expression of the harmonic coefficients, that can be expressed in the polar complex plane, with $z = \rho_s e^{i \theta_s}$, as
\begin{equation}
	\label{deqn_coeff}
	C_n= -\frac{\mu_0 I}{2 \pi} \frac{R_{ref}^{n-1}}{z_s^n} = -\frac{\mu_0 I}{2 \pi} \frac{R_{ref}^{n-1}}{\rho_s^n} e^{-i n \theta_s}
	\
\end{equation}

\section{Symmetric cross section}
\label{sec:symmetric}
As it has been shown for the \textit{asymmetric} configuration, we will first check if there are solutions for the symmetric configuration by considering the following constraints:

\begin{itemize}
	\item All current lines are perpendicular to the Cartesian $xy$-plane along the straight section of the magnet.
	\item The magnitude of the current is the same for all conductors, where only its sign changes. The current can be therefore expressed as 
	\begin{equation*}
		\label{deqn_current3}
		I_j=  s_j I \quad\textrm{with}\quad s_j \in \{-1,1\}
		\
	\end{equation*}
	In this regard, the current sign, $s_j$, is considered as to create a positive $B_y$ and therefore $I$ is negative on the \textit{right} (i.e. $x>0$) and positive on the \textit{left} (i.e. $x<0$).
	\item All current lines are equidistant to the origin $z=0$, at a distance $\rho$.
	\item The multipole expansion will be considered within a circular domain $D$ centered at $z=0$, and with a radius $R_{ref}$, related to the magnet apeture $\rho$ as $R_{ref} = 2/3 \rho$.
	\item An odd number of current lines, $m_r$ are considered on the \textit{right} (i.e. $x>0$) and the \textit{left }(i.e. $x<0$), with a total of current lines $m=2m_r$.
	\item \textit{Top-bottom} and \textit{left-right} symmetries are considered, therefore cancelling all skew components of the harmonic coefficients.
	\item From $C_{2}$ all harmonic coefficients must be $0$ up to $C_{n_0}$:
	\begin{equation*}
		\label{deqn_current}
		C_{n,total}= 0  \quad\textrm{with}\quad n \in \{2,3,\dots,n_0\}
		\
	\end{equation*}
	In this case, due to the geometrical configuration, all even coefficients are automatically canceled.
	
	\item The minimum angular distance between adjacent current lines is $\theta_{min}$.
	
\end{itemize}

In the symmetric case, given the higher degree of symmetry, the number of components of $\theta$ is further reduced. For an odd number of current lines on each side (i.e. \textit{left} and \textit{right}), the only way to maintain a \textit{top-bottom} symmetry is to place a conductor in the mid-plane (i.e. $y=0$). Therefore, for a total of $m$ current lines to form the magnet, where $m=2m_r$, the total number, $m_{opt}$, of argument variables, $\theta_i$, to characterize the full position of all conductors is $m_{opt} = (m-2)/4$, for a given $\rho$.

The array $\theta$ of size $m_{opt}$ where the argument of the conductors will be stored can be expressed as
\begin{equation*}
	\label{deqn_theta_array}
	\theta = \left\{  \theta_1,\theta_2,\dots \theta_{m_{opt}} \right\}
\end{equation*}
We can now write an example of a general equation $f_{min}$ to be minimized
\begin{equation}
	\label{deqn_fmin}
	f_{min} = -2  - 4\sum_{j=1}^{m_{opt}} \cos{\theta_j}
\end{equation}

In an analogous form, one can find the set of equality constraints to cancel specific harmonic coefficients. The general expression to cancel harmonic coefficients for a symmetric configuration is
\begin{equation}
	\label{deqn_sym_coef}
	\left( -1+e^{in\pi}   \right) \left(1 + 2 \sum_{j=1}^{m_{opt}} \cos{n\theta_j}\right) = 0
\end{equation}

In this example, as for the asymmetric case explored before, a minimum angular spacing between conductors was included as
\begin{equation}
	\label{deqn_spacing_sym}
	\min \left\{  \| \theta_j - \theta_{j+1} \| \right\} - \theta_{min} \geq 0, \quad j = 1,\dots,m_{opt}-1 \\
\end{equation}

The overall optimization equations can therefore be written for the symmetric form as

\begin{equation}
	\label{deqn_opti_sy}
	\eqalign{\textrm{minimize}  \left( -\sum_{j=1}^{m_{opt}} \cos{\theta_j} \right) \cr
		\textrm{subject to:}  \cr
		\min \left\{  \| \theta_j - \theta_{j+1}   \|    \right\} - \theta_{min} \geq 0, \quad j= 1,\dots,m_{opt}-1  \cr
		\theta_{min} \leq \theta_j  \leq \frac{\pi}{2}, \quad j = 1, \dots,m_{opt}/2 \cr
		\frac{\pi}{2}\leq \theta_j  \leq \pi, \quad j = m_{opt}/2+1, \dots,m_{opt}  \cr
		\left( -1+e^{in\pi}   \right) \left(1 + 2 \sum_{j=1}^{m_{opt}} \cos{n\theta_j}\right)  = 0, \quad n = 3,5, \dots,n_0
	} 
\end{equation}

An example of a solution to (\ref{deqn_opti_sy}) is illustrated in Figure~\ref{fig_UL}(b). Analogous equations can also be formulated to create higher order symmetric magnets, e.g. quadrupoles, sextupoles.

\section{Uni-layer magnet's ends geometry}
\label{sec:magnet_end_geometry}

\subsection{Radius of curvature of the magnet's ends}

From the generalized superelliptical space curve wrapped around a cylindrical surface $\mathbf{\gamma_{gse}}$,  equation~(\ref{deqn_gsellipse}), one can derive its radius of curvature a $R_{gse}$ from its curvature $\kappa_{gse}$ as
\begin{equation}
	\label{deqn_R_gse}
	R_{gse} = \frac{1}{\kappa_{gse}} = \frac{\|\mathbf{\gamma_{gse}}'\|^3}{\|\mathbf{\gamma_{gse}}' \times \mathbf{\gamma_{gse}}''\|} \\ 
\end{equation}
where the \textit{prime} symbol designates differentiation with respect to the parametrization variable, $\Phi$ in the case of $\mathbf{\gamma_{gse}}$. The solution to equation (\ref{deqn_R_gse}) can be expressed as

\begin{equation}
	\label{deqn_R_gse_solution}
	\eqalign{
	R_{gse} &= \\
	& \frac{\left(\chi^{2} l_{t}^{2} \cos^{\frac{4}{\zeta}}{\left(\Phi \right)} \tan^{2}{\left(\Phi \right)} + \frac{\delta_{i}^{2} \rho^{2} \zeta^{2} \sin^{\frac{4}{\chi}}{\left(\Phi \right)}}{\tan^{2}{\left(\Phi \right)}}\right)^{\frac{3}{2}}}{\delta_{i} \rho \zeta \sin^{-2}{\Phi} }  \\	
	& \cdot \biggl[ \chi^{2} l_{t}^{2} \biggl( \chi^{2} \zeta^{2} - 2 \chi^{2} \zeta \sin^{2}{\left(\Phi \right)} + \chi^{2} \sin^{4}{\left(\Phi \right)} \\
	&+ 2 \chi \zeta^{2} \sin^{2}{\left(\Phi \right)} - 2 \chi \zeta^{2} - 2 \chi \zeta \sin^{4}{\left(\Phi \right)} \\
	&+ 2 \chi \zeta \sin^{2}{\left(\Phi \right)} + \delta_{i}^{2} \zeta^{2} \sin^{\frac{4}{\chi}}{\left(\Phi \right)} \\
	&- 2 \delta_{i}^{2} \zeta^{2} \sin^{2 + \frac{4}{\chi}}{\left(\Phi \right)} \\
	& + \delta_{i}^{2} \zeta^{2} \sin^{4 + \frac{4}{\chi}}{\left(\Phi \right)}+ \zeta^{2} \sin^{4}{\left(\Phi \right)} \\
	&- 2 \zeta^{2} \sin^{2}{\left(\Phi \right)} + \zeta^{2} \biggr) \sin^{2 + \frac{4}{\chi}}{\left(\Phi \right)} \cos^{-2 + \frac{4}{\zeta}}{\left(\Phi \right)} \\
	&+ \delta_{i}^{4} \rho^{2} \zeta^{4} \sin^{-2 + \frac{12}{\chi}}{\left(\Phi \right)} \cos^{6}{\left(\Phi \right)} \biggr]^{-\frac{1}{2}}
	}
\end{equation}

In an analogous form, we can also derive the radius of curvature of the superelliptical space curve $\mathbf{\gamma_{se}}$, equation~(\ref{deqn_sellipse}), where the solution can be expressed as
\begin{equation}
	\label{deqn_R_se}
	\eqalign{
	R_{se} &= \frac{1}{\kappa_{se}} = \frac{\|\mathbf{\gamma_{se}}'\|^3}{\|\mathbf{\gamma_{se}}' \times \mathbf{\gamma_{se}}''\|} \\
	&=\frac{\left(\frac{4 b^{2} \sin^{\frac{4}{\zeta}}{\left(\varphi \right)}}{\tan^{2}{\left(\varphi \right)}} + \delta_{\theta}^{2} \rho^{2} \cos^{\frac{4}{\zeta}}{\left(\varphi \right)} \tan^{2}{\left(\varphi \right)}\right)^{\frac{3}{2}} }{ \delta_{\theta} \rho \cos^{-2}{\left(\varphi \right)}} \\
	& \cdot \biggl[ 4 b^{2} \biggl( \delta_{\theta}^{2} \sin^{4}{\left(\varphi \right)} \cos^{\frac{4}{\zeta}}{\left(\varphi \right)} + 4 \zeta^{2} \\ 
	&- 8 \zeta + 4 \biggr)  \sin^{-2 + \frac{4}{\zeta}}{\left(\varphi \right)} \cos^{2 + \frac{4}{\zeta}}{\left(\varphi \right)} \\
	& + \delta_{\theta}^{4} \rho^{2} \sin^{6}{\left(\varphi \right)} \cos^{-2 + \frac{12}{\zeta}}{\left(\varphi \right)} \biggr]^{-\frac{1}{2}}}
\end{equation}

\subsection{Length of the curves}
The differential element $ds$ of the space curve $\mathbf{\gamma_{gse}}$ can be expressed as

\begin{equation}
	\label{deqn_ds_se}
	\eqalign{
		ds &= \sqrt{dx^2+dy^2+dz^2} = \|\mathbf{\gamma_{gse}}'\| \\
		&= 2 \sqrt{\frac{l_{t}^{2} \cos^{\frac{4}{\zeta}}{\left(\Phi \right)} \tan^{2}{\left(\Phi \right)}}{\zeta^{2}} + \frac{\delta_{i}^{2} \rho^{2} \sin^{\frac{4}{\chi}}{\left(\Phi \right)}}{\chi^{2} \tan^{2}{\left(\Phi \right)}}}}
\end{equation}

The length of the curve $l_{gse}$ can be therefore expressed as
\begin{equation}
	\label{deqn_length_gse}
	\eqalign{
	l_{gse} &= \\
	&2 \int_{0}^{\frac{\pi}{2}} \sqrt{\frac{l_{t}^{2} \cos^{\frac{4}{\zeta}}{\left(\Phi \right)} \tan^{2}{\left(\Phi \right)}}{\zeta^{2}} + \frac{\delta_{i}^{2} \rho^{2} \sin^{\frac{4}{\chi}}{\left(\Phi \right)}}{\chi^{2} \tan^{2}{\left(\Phi \right)}}} d\Phi}
\end{equation}
which is an integral that can be evaluated numerically.

The analogous differential element $ds$ of the space curve $\mathbf{\gamma_{se}}$ can be expressed as
\begin{equation}
	\label{deqn_ds_se}
	\eqalign{
	ds &= \sqrt{dx^2+dy^2+dz^2} = \|\mathbf{\gamma_{se}}'\| \\
	&= \frac{\sqrt{\frac{4 b^{2} \sin^{\frac{4}{\zeta}}{\left(\varphi \right)}}{\tan^{2}{\left(\varphi \right)}} + \delta_{\theta}^{2} \rho^{2} \cos^{\frac{4}{\zeta}}{\left(\varphi \right)} \tan^{2}{\left(\varphi \right)}}}{\zeta}}
\end{equation}

The length of the curve $l_{se}$ can be therefore expressed as
\begin{equation}
	\label{deqn_length_ge}
	l_{ge}= \int_{0}^{\frac{\pi}{2}} \frac{\sqrt{\frac{4 b^{2} \sin^{\frac{4}{\zeta}}{\left(\varphi \right)}}{\tan^{2}{\left(\varphi \right)}} + \delta_{\theta}^{2} \rho^{2} \cos^{\frac{4}{\zeta}}{\left(\varphi \right)} \tan^{2}{\left(\varphi \right)}}}{\zeta} d\varphi \\
\end{equation}
which is also an integral that can be evaluated numerically.

\section{Canted cosine theta dipole magnets}
\label{sec:CCT}

The general equation of the space curve $\mathbf{\gamma_{CCT}}$ describing a canted cosine theta (CCT) dipole lying in a cylindrical surface of radius $\rho$, can be expressed in Cartesian coordinates as

\begin{equation}
	\label{deqn_CCT}
	\eqalign{
	\mathbf{\gamma_{CCT}}(\phi) &= \rho \cos{\left(\phi \right)}\mathbf{\hat{i}} + \rho \sin{\left(\phi \right)}\mathbf{\hat{j}} \\
	& + \left(\frac{\omega \phi}{2 \pi} + \frac{\rho \sin{\left(\phi \right)}}{\tan{\left(\alpha \right)}}\right)\mathbf{\hat{k}}\\}
\end{equation}
where $\phi$ is the parametrization variable, $\omega$ is axial distance between adjacent turns, and $\alpha$ is the inclination.

The radius of curvature $R_{CCT}$ of the space curve $\mathbf{\gamma_{CCT}}$ can also be derived as it has been done in (\ref{deqn_R_gse}) and (\ref{deqn_R_se}) as
\begin{equation}
	\label{deqn_R_cct}
	\eqalign{
		R_{CCT}  &=  \frac{1}{\kappa_{CCT}}  \\
		&= \frac{\|\boldsymbol{\gamma_{CCT}}'\|^3}{\|\boldsymbol{\gamma_{CCT}}' \times \boldsymbol{\gamma_{CCT}}''\|}  \\
		&= \left(4 \pi^{2} \rho \tan^{2}{\left(\alpha \right)} \right)^{-1}\\
		& \cdot  \biggl(4 \pi^{2} \rho^{2} \tan^{2}{\left(\alpha \right)} \\
		& + \left(\omega \tan{\left(\alpha \right)} + 2 \pi \rho \cos{\left(\phi \right)}\right)^{2}  \biggr)^{\frac{3}{2}}\\
		& \cdot \biggl( - \omega^{2} + \frac{\omega^{2}}{\cos^{2}{\left(\alpha \right)}} \\
		&+ 4 \pi \omega \rho \cos{\left(\phi \right)} \tan{\left(\alpha \right)} + \frac{4 \pi^{2} \rho^{2}}{\cos^{2}{\left(\alpha \right)}} \biggr)^{-\frac{1}{2}}   }
\end{equation}

The minimum radius of curvature of $\mathbf{\gamma_{CCT}}$ can be found periodically at $\phi =  \pi/2 + n2\pi \; \textrm{and} \;3\pi/2 + n2\pi $. We can therefore simplify (\ref{deqn_R_cct}), leading to the equation for the minimum radius of curvature along $\mathbf{\gamma_{CCT}}$ 

	\begin{equation}
	\label{deqn_R_cct_min}
	\eqalign{
	R_{CCT,min} &=  \frac{\left(\omega^{2} + 4 \pi^{2} \rho^{2}\right)^{\frac{3}{2}} {\sin{\left(\alpha \right)}}}{4 \pi^{2} \rho \sqrt{\omega^{2} \sin^{2}{\left(\alpha \right)} + 4 \pi^{2} \rho^{2}}} \\
	&\quad\textrm{with}\quad \alpha \in \left(0,\pi/2 \right)\\}
\end{equation}

	The \textit{maximum }of the \textit{minimum} radius of curvature can be found as we get closer to the limit of the boundary of the interval of $\alpha$
	\begin{equation*}
		\label{deqn_R_cct_min}
		\lim_{\alpha \to \pi/2} R_{CCT,min} = \rho + \frac{\omega^{2}}{4 \pi^{2} \rho} \\
	\end{equation*}
	
	The value of $R_{CCT,min}$ is mainly determined by the angle $\alpha$, since for real CCT magnets $\rho$ is at least an order of magnitude higher than $\omega$.
	
	The length of the conductor used for a given straight section can be calculated based on the length of an individual turn \cite{Brouwer2015}. The differential element $ds$ of the space curve $\mathbf{\gamma_{CCT}}$ can be expressed as
	
	\begin{equation}
		\label{deqn_ds}
		\eqalign{
		ds &= \sqrt{dx^2+dy^2+dz^2} = \|\mathbf{\gamma_{CCT}}'\| \\
		&= \rho \sqrt{1 + \left(\frac{\omega}{2 \pi \rho} + \frac{\cos{\left(\phi \right)}}{\tan{\left(\alpha \right)}}\right)^{2}}\\}
	\end{equation}
	
	The length of the turn $l_{CCT}$ can be therefore expressed as
	\begin{equation}
		\label{deqn_length_CCT}
		l_{CCT}= \int_{0}^{2\pi} \rho \sqrt{1 + \left(\frac{\omega}{2 \pi \rho} + \frac{\cos{\left(\phi \right)}}{\tan{\left(\alpha \right)}}\right)^{2}}d\phi \\
	\end{equation}
	which can be computed numerically.
	
	In the case of a real magnet, and considering a cable with a thickness $t_{CCT}$, the thickness of the rib $t_{rib}$ separating consecutive turns at the mid-plane is
	
	\begin{equation}
		\label{thickness_rib}
		t_{rib}= \omega \sin{\alpha}  - t_{CCT} \\
	\end{equation}
	
	The total length of conductor used in the CCT magnet along certain straight section $L_{ss}$ can be computed \cite{Brouwer2015} as 
	
	\begin{equation}
		\label{deqn_total_length_CCT}
		L_{CCT,T}=  \frac{L_{ss} }{\omega} l_{CCT} \\
	\end{equation}
	where $ l_{CCT}$ is defined in (\ref{deqn_length_CCT}). One can then compute the total volume of conductor by assuming certain dimensions for the cross section of the conductor.
	
	The $B_{m,CCT}$ bore dipole field created by a single layer \cite{Brouwer2015}, considering $\rho \gg \omega$, can be expressed as 
	
	\begin{equation}
		\label{deqn_B_CCT}
		B_{y,CCT}=  \frac{- \mu_0 I }{2 \omega \tan{\alpha}}\\
	\end{equation}

\section{CT sector coil}
\label{sec:CT}
Let's consider the simplest form of a $\cos{\theta}$ type dipole magnet, where we will analyze a sector coil constructed with a conductor that carries a uniform current density $j_{eng}$, where $dI =j_{eng} \rho d\rho d\theta$,  we can derive the main dipole field $B_y$ from (\ref{deqn_line}) 

\begin{equation}
	\label{deqn_sector}
	B_y = \textrm{Re} \left\{ \mathbf{B}(z) \right\}  =   \textrm{Re} \left\{-\frac{\mu_0 I}{2 \pi} \frac{1}{z_s}\right\} = -\frac{\mu_0 I}{2 \pi} \frac{\cos{\theta}}{\|z_s\|}\\
\end{equation}
leading to the differential expression of $B_y$ based on a uniform current density
\begin{equation}
	\label{deqn_d_sector}
	dB_y = -\frac{\mu_0 j_{eng} \cos{\theta} }{2 \pi} d\rho d\theta\\
\end{equation}

The overall field can be found by integrating (\ref{deqn_d_sector}) for sector coils defined by its angle $\beta$, width $w_c$, and inner radius $r_i$ \cite{Rossi2007} as
\begin{equation}
	\label{deqn_b_sector}
	\eqalign{
	B_y &= -\frac{\mu_0 j_{eng} }{\pi} \int_{-\beta}^{\beta} \int_{r_i}^{r_i+w_c}\cos{\theta} d\rho d\theta \\
	& = -\frac{2\mu_0 j_{eng} }{\pi} w_c \sin{\beta}\\}
\end{equation}

The area of a sector coil of angle $\beta$ can be written as
\begin{equation*}
	\label{deqn_area_CS}
	A_{sector}=  2 \beta w_c \left( 2\rho + w_c\right)  \\
\end{equation*}

\section{Comparison between UL and CCT concepts for equal amount of conductor}
\label{sec:comparison_UL}
The field generated along the straight section of a UL magnet can be computed based on the principle of superposition and equation~\ref{deqn_line}. The total length of conductor required per unit of length of straight section of the magnet is proportional to the total number of current lines in its cross section $m$, and is independent of the radius of the aperture of the magnet $\rho$.

If we rearrange the same length of conductor into a CCT, given the parameters $\alpha$ and $\rho$, one could compute the parameter $\omega$ that will be required. This problem can be numerically solved based on equation~\ref{deqn_total_length_CCT}.
The field generated by a single layer with the computed $\omega$ and given $\alpha$ and $\rho$ can be computed based on equation~\ref{deqn_B_CCT}.

\section{Approximation for a comparison between magnet concepts using the CT sector coil}
\label{sec:comparison}
The field generated along the straight section is compared between CT, SMCT and UL (considered equivalent in this approximation) in relation to the field generated by the CCT. The following points are considered:

\begin{itemize}
	\item Structural elements within SMCT, CCT and UL are disregarded.
	\item The field generated by CT, SMCT and UL is considered equivalent since a very similar cable layout within the straight section is possible.
	\item A constant current density $j_{eng}$ is considered for all magnets.
	\item An equal volume of conductor $V_{cond}$ is considered for all magnets.
	\item In the case of CT, SMCT and UL concepts, the generated dipole field can be approximated by a $60^{\circ}$ CT sector coil \cite{Rossi2007} , equation~\ref{deqn_b_sector}.

\end{itemize}

The assumptions for the geometrical parameters of the CCT are summarized as:
\begin{itemize}
	\item The diameter of the aperture of the CCT layer is $2\rho$
	\item The thickness of the cable: $t_{cable} = \rho/20$
	\item The thickness of the rib: $t_{rib} = \rho/200$
	\item The width of the cable: $w_{cable} = \rho/2$
\end{itemize}

\begin{figure}[!h]
	\centering
	\includegraphics[width=3.275in]{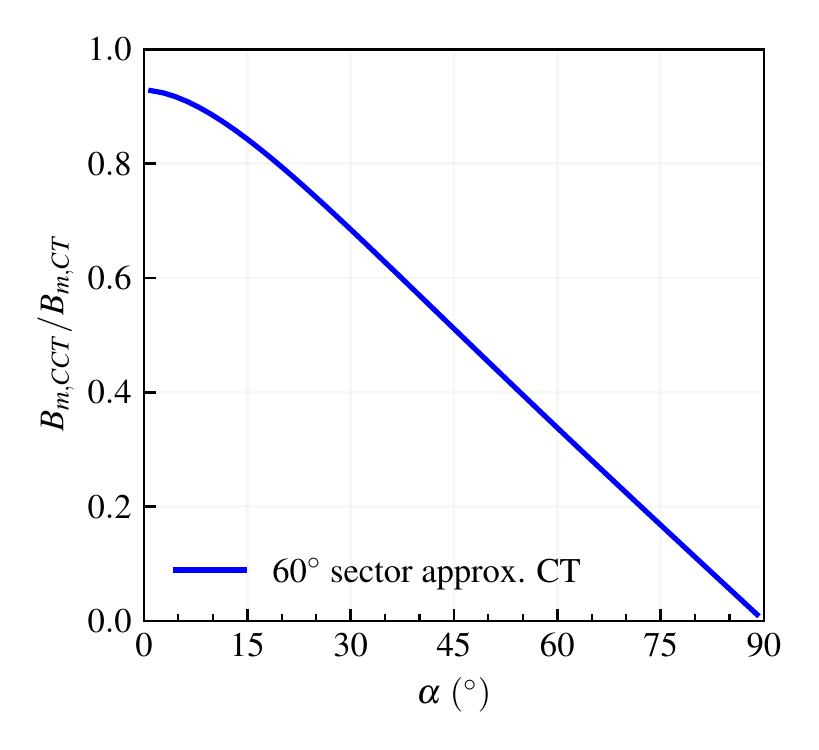}
	\caption{Ratio between the main field generated by CCT magnet $B_{m,CCT}$ and the CT sector approximation $B_{m,CT}$  (equivalent for CT, SMCT and UL layers) .}
	\label{fig_ratio_field}
\end{figure}

From the above assumptions, one can compute $B_{m,CCT}$ (from equation~\ref{deqn_B_CCT}) and the volume of conductor $V_{cond}$ used for a given straight section length (based on equation~\ref{deqn_total_length_CCT}). At the same time, one can compute an equivalent volume of conductor for a CT sector coil, that will result in a given coil width, $w_c$, and that will generate a specific $B_{m,CT}$ (based on equation~\ref{deqn_b_sector}). The ratio between the two fields as a function of the inclination angle $\alpha$ is illustrated in Figure~\ref{fig_ratio_field}.

\newpage
\bibliographystyle{iopart-num}
\bibliography{paper.bib}

\end{document}